\documentclass[preprint,12pt, a4paper]{elsarticle}

\usepackage{amssymb}
\usepackage{soul}
\usepackage{placeins}

\usepackage{tikz}
\usetikzlibrary{positioning}
\usetikzlibrary{shapes,snakes}

\usepackage{listings}
\usepackage{color}

\definecolor{mygreen}{rgb}{0,0.6,0}
\definecolor{mygray}{rgb}{0.5,0.5,0.5}
\definecolor{mymauve}{rgb}{0.58,0,0.82}

\usepackage{siunitx}
\usepackage{multirow}

\usepackage{hyperref}

\usepackage{float}
\restylefloat{table}

\journal{SoftwareX}

\begin{document}
\lstset{language=Python}

\begin{frontmatter}



\title{PyfastSPM: A Python package to convert 1D FastSPM data streams into publication quality movies}


\author[label1]{K.\ Briegel\corref{contrib}}
\author[label1]{F.\ Riccius\corref{contrib}}
\cortext[contrib]{These authors contributed equally.}
\author[label1]{J.\ Filser}
\author[label1]{A.\ Bourgund}
\author[label1]{R.\ Spitzenpfeil}
\author[label2]{M.\ Panighel}
\author[label2,label3]{C.\ Dri\corref{presaddr}}
\cortext[presaddr]{Present address: cpi-eng S.r.l.\ - via del Lavatoio, 4 - 34132 Trieste (Italy)}
\author[label4]{B.A.J.\ Lechner\corref{cor1}}
\ead{bajlechner@tum.de}
\cortext[cor1]{Corresponding author.}
\author[label1]{F.\ Esch}

\address[label1]{Chair of Physical Chemistry, Department of Chemistry \& Catalysis Research Center, Technical University of Munich, D-85748 Garching, Germany}
\address[label2]{CNR-IOM Laboratorio TASC, S.S. 14 km 163.5, Basovizza, I-34149 Trieste, Italy}
\address[label3]{Elettra-Sincrotrone Trieste, S.S. 14 km 163.5, Basovizza, I-34149 Trieste, Italy}
\address[label4]{Functional Nanomaterials, Department of Chemistry \& Catalysis Research Center, Technical University of Munich, D-85748 Garching, Germany}

\begin{abstract}
In a continued quest to monitor subsecond surface dynamics on the atomic scale and to improve imaging resolution, a FAST module to accelerate existing scanning probe microscopy setups was previously presented. Hereby, the speedup is enabled by external electronics without modification of the actual instrument. The resulting one-dimensional (1D) data stream, recorded while the tip oscillates in a sinusoidal motion, has to be reconstructed into a layered rectangular matrix in order to visualize the movie. The Python-based \texttt{pyfastspm} package performs this conversion, while also correcting for sample tilt, noise frequencies, piezo creep, and thermal drift. Quick automatic conversion even of considerable batches of data is achieved by efficient algorithms that bundle time-expensive steps, such as interpolation based on Delaunay triangulation.

\end{abstract}


\end{frontmatter}

\section*{Code Metadata}

% \section*{Required Metadata}
% 

% \section*{Current code version}
% 

\begin{table}[H]
\begin{tabular}{|l|p{6.5cm}|p{6.5cm}|}
\hline
\textbf{Nr.} & \textbf{Code metadata description} & \textbf{Please fill in this column} \\
\hline
C1 & Current code version & v1.0.1 \\
\hline
C2 & Permanent link to code/repository used for this code version & \url{https://doi.org/10.5281/zenodo.6824215} \\
\hline
C4 & Legal Code License   & MIT, Apache-2.0 \\
\hline
C5 & Code versioning system used & git \\
\hline
C6 & Software code languages, tools, and services used & Python 3, FFmpeg \\
\hline
C7 & Compilation requirements, operating environments \& dependencies & \url{http://fastspm.gitlab.io/pyfastspm} \\
\hline
C8 & If available Link to developer documentation/manual & \url{http://fastspm.gitlab.io/pyfastspm/} \\
\hline
C9 & Support email for questions & carlo.dri@gmail.com\\
\hline
\end{tabular}
% \caption{Code metadata}
\label{} 
\end{table}

%\linenumbers


% The permanent link to code/repository or the zip archive should include the following requirements: 

% README.txt and LICENSE.txt.

% Source code in a src/ directory, not the root of the repository.

% Tag corresponding with the version of the software that is reviewed.

% Documentation in the repository in a docs/ directory, and/or READMEs, as appropriate.

\section{Motivation and significance}

% Introduce the scientific background and the motivation for developing the software.

% Explain why the software is important, and describe the exact (scientific) problem(s) it solves.

% Indicate in what way the software has contributed (or how it will contribute in the future) to the process of scientific discovery; if available, this is to be supported by citing a research paper using the software.

% Provide a description of the experimental setting (how does the user use the software?).

% Introduce related work in literature (cite or list algorithms used, other software etc.).

On the sub-second timescale, a multitude of surface dynamics occur that can be revealed particularly well by video-rate scanning probe microscopies (SPM) -- from equilibrium molecular diffusion \cite{henss2019density, bourgund2019influence} to film growth \cite{patera2018real}, from cluster isomerization \cite{lechner2018microscopy} to potential-dependent electrochemical surface processes \cite{wei2021electrochemical, rost2018high}. Time-resolved SPM techniques can reveal particular dynamics at the atomic scale that would otherwise be overlooked \cite{patera2018real}. 

In SPM, a point-like sensor (tip) typically moves stepwise across a sample and records local information on a square spatial grid. In scanning tunnelling microscopy (STM), for example, topographic and chemical information is obtained from the tunnelling current flowing between the tip and the sample. This current is typically kept constant by a feedback loop that regulates the vertical tip position, which is in fact the measured quantity. Stepwise motion and feedback make the measurement inherently slow.

While several successful, yet experimentally demanding and expensive implementations of fast SPM setups allow for imaging under full feedback conditions \cite{rost2018high, schitter2008scanning}, most experimental groups face the challenge of accelerating their existing, slow setups to access fast surface dynamics. This speed-up can be realized with the previously reported FAST module by simply inserting an electronics add-on module that performs a synchronized fast scanning and tunnelling current acquisition \cite{dri2019new}. It comes at the price, however, of quasi-constant height imaging, where the current is measured, while the feedback is slowed down such that it only controls the average tip position. Moreover, a continuous sinusoidal oscillation of the tip in the fast scan direction replaces the triangular stepwise motion. Data points now are recorded continuously, by binning over constant-duration time steps. Hence, lateral positions corresponding to each bin have to be determined retrospectively during data evaluation, by means of adequate interpolation, which is the central topic of this paper. Similar to recent spiral movement approaches \cite{Junkes:ICALEPCS2017-THPHA154}, a dedicated image reconstruction software is required to create movies and image frames that reveal the dynamics of interest for further data analysis.

Here, we present a Python-based software package, called \texttt{pyfastspm}, which is openly accessible and easy to use for a non-dedicated user. It converts 1D SPM data streams, such as those recorded by the FAST module, into 2D MPEG4 movies, optionally in batch automation. Specifically, the software corrects for several aspects relating to the tip movement across the surface, as well as for additional interfering signals:
\begin{enumerate}
    \item \textit{Background frequencies}. Residual sample tilt, which occurs mainly in the fast $x$-direction, can be removed by filtering out $x$ and $y$ scanning frequencies and their overtones in the original 1D data time series via FFT. In this process, additional background noise, e.g. of electronic or mechanical origin, can be eliminated as well.
    \item \textit{Tip path}. Common movie formats require the pixels to be placed on square grids. Thus, the complex tip path has to be taken into account correctly while transforming the acquired 1D SPM data stream into a 3D stack consisting of 2D movie frames. This correction occurs on the basis of an estimated tip path that takes into account (a) the sinusoidal tip motion in the fast $x$-direction, including phase shifts between tip motion and data acquisition; (b) the triangular tip motion in the slow $y$-direction; and (c) the so-called creep distortions due to acceleration delays at the tip turnaround points between frames. It is performed based on a Delaunay transformation \cite{computational_geometry}, with parameters that can be automatically extracted from the FastSPM data stream. As a result, every single measured pixel can be included in the movie (from forward and backward, up and down motion), allowing a doubling of both, the temporal and spatial movie resolution. 
    \item \textit{Lateral drift}.  At the atomic scale, lateral image shifts are often encountered due to thermal drift. The \texttt{pyfastspm} package determines and corrects for the lateral drift path by evaluating image correlations between entire movie frames at variable temporal delays, after the above corrections have been made. Averaging over frames of a drift-corrected movie can lead to superior spatial resolution of structures underlying highly mobile, streaky surface dynamics \cite{arndt2020order}.
\end{enumerate}

In the following, we will describe these steps in more detail, in the sequence in which they are called in a typical movie reconstruction and conversion pipeline. \texttt{pyfastspm} unleashes the full potential of SPM when it comes to exploring surface dynamics studies \cite{bourgund2019influence, patera2018real, lechner2018microscopy, arndt2020order} and gives the experimentalist the freedom to focus on the scientific questions at hand. In the following, we give STM examples in which the acquired signal is tunneling current, but \texttt{pyfastspm} is generally applicable for any kind of local probe microscopy.

\section{Software description}

\texttt{pyfastspm} is a standalone Python package, but also comes with a Jupyter notebook that includes standard input parameters and default values for easy conversion.

\subsection{Software Architecture}

\begin{figure}[H]
    \centering
    \includegraphics[width=1\textwidth]{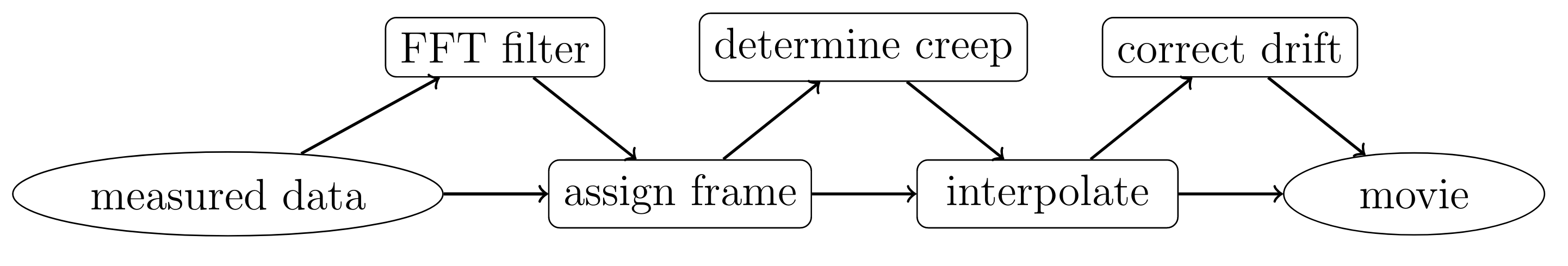}
    \caption{The workflow sequence  of PyfastSPM functionalities, including the two main transformation steps \textit{assign frame} and \textit{interpolate} as well as optional corrections, to create a movie from the measured 1D data stream.}
    \label{fig:sketch}
\end{figure}

Central to the package is the \texttt{FastMovie} class that loads the acquired 1D SPM data stream $I(t)$ from a Hierarchical Data Format \texttt{HDF5} file. Fig. \ref{fig:sketch} shows the general workflow, including all transformations to the final 3D stack of 2D frames $I(x,y,t)$ that can be exported as a movie. The first step, \textit{assign frame}, sections the data stream into the respective frames. The second main step removes the distortion from the measured sine-shaped tip path by \textit{interpolating} onto a square grid. These are the minimum steps required to write the data into a movie file. Further optional correction steps (upper row in the Figure) can be included to remove measurement artefacts: \textit{FFT filter} removes background frequencies in the time domain, \textit{determine creep} identifies additional distortions at the tip turnaround point in the $y$-direction and \textit{correct drift} shifts frames with respect to each other according to their lateral drift.

\subsection{Software Functionalities}

\paragraph{FFT filter}
This function applies low-, high- and band-pass filters of selectable frequencies and widths to the FFT-transformed 1D data stream. Furthermore, the convolution with a Gaussian of selectable width can be applied to eliminate noise in the fast $x$-direction. Back transformation results in background-corrected and smoothed data that typically do not require further filtering in the spatial domain.

\begin{figure}[htbp]
    \centering
    \includegraphics[width=1\textwidth]{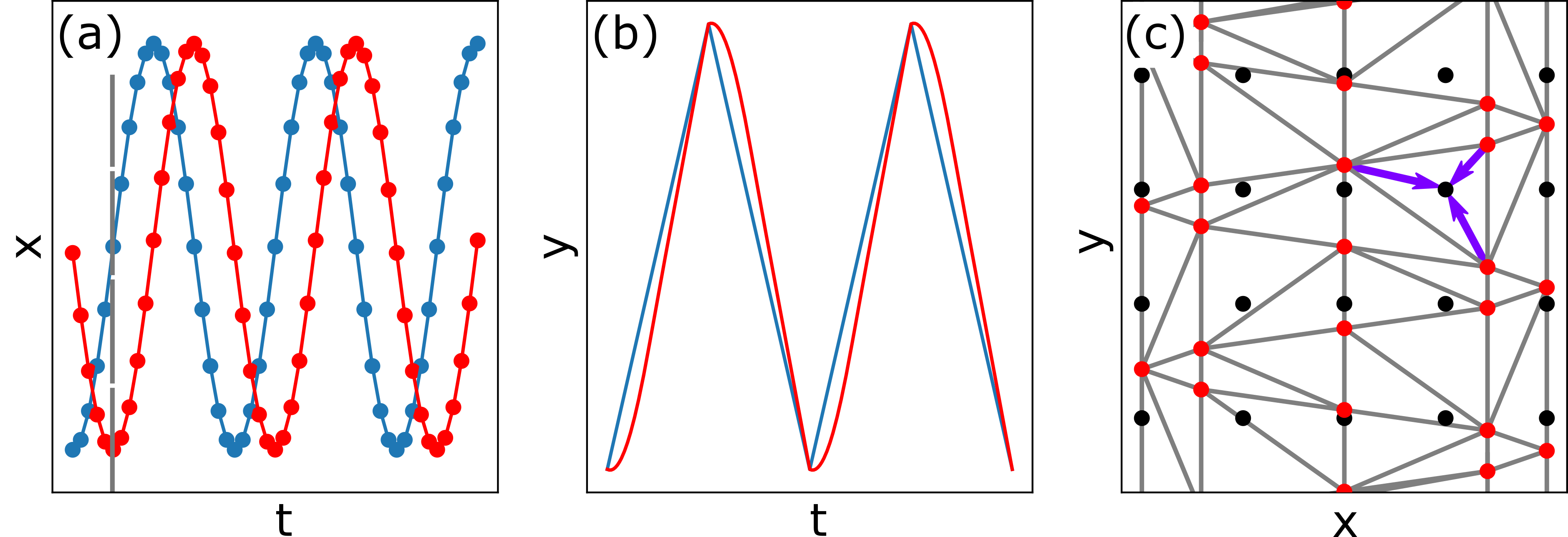}
    \caption{Tip path in (a) x- and (b) y-directions as a function of time (solid lines). The nominal tip path is indicated in blue, the actual one in red. The deviations stem from a delayed piezo response and creep, respectively. Dots in (a) mark measurement positions along x. Note that the image reconstruction always starts at an x-turnaround point (gray dashed line). (c) The deduced x/y tip positions, indicated by the red dots, are then mapped onto a square grid, marked in black. The Delaunay triangulation is indicated by the gray lines. The resulting barycentric coordinates used for interpolation are exemplified by the purple arrows.}
    \label{fig:schematic}
\end{figure}

\paragraph{assign frame}
A complete period of the tip movement results in an \textit{image} that is composed of an up 'u' and down 'd' \textit{frame} pair (slow $y$-movement), each composed of forward and backward lines (fast $x$-movement). This conversion step distributes the data points into frames. \\
To account for the overall delay between the physical tip position and consequently the acquired data in the forward and backward lines, a phase shift in the fast $x$-direction needs to be applied, as illustrated by the blue and red curves in Fig. \ref{fig:schematic}(a), respectively. This phase shift is calculated automatically by correlating the current signal of sequential forward and backward lines or defined manually by the user. A well-corrected phase allows the combination of forward 'f' and backward 'b' parts of the frames into a single, interlaced 'i' frame. The resulting movie mode that contains the full acquired data is called 'udi', but also movies with partial data sets can be generated, e.g.\ 'uf', 'udb', etc. An 'i' frame contains twice the data points of a 'f' or a 'b' frame. A 'ud' movie contains twice the frames of a 'u' or a 'd' movie.

\paragraph{determine creep}
While the sinusoidal fast $x$-direction has a smooth turnaround to allow the piezo to perfectly follow the intended motion, the sharp turnaround in the slow $y$-direction leads to a significant piezo delay (creep). Other approaches avoid such creep problems in the experiment by suppressing high frequency Fourier components \cite{rost2005scanning}, while here we add an optional post-measurement correction to \texttt{pyfastspm}. The piezo creep leads to a distorted movement in $y$ that is illustrated in red in Fig. \ref{fig:schematic}(b), resulting in a compression at the bottom of the up frame and at the top of the down frame. In a single frame, this is often barely perceptible, but in 'ud' movies, it results in a flickering (see left side of the Supporting Movie).\\
Here, creep is corrected for by distorting sequential up and down frames in $y$ until they match. This distortion is calculated on a selected line of pixels in $y$-direction by fitting the $y$-positions with a B-spline as a computationally cheap curve fitting approach. In order to account for possible differences between microscopes, three functions have been implemented in \texttt{pyfastspm} to map the $y$-positions, a B\'{e}zier function, a sine function that merges into a linear function, and a square root correction to a linear function (as described by Choi and coworkers \cite{choi2014growth}). The B\'{e}zier fit is less constrained and thus maps the creep well and relatively quickly, but it can fail in the absence of pronounced features. Here, the more constrained sine and square root options yield a more stable fit; in our experience, the sine function yields the best results for batch conversions (see Supporting Movie). Finally, the user has the option to manually input known creep parameters, e.g. those obtained from measurements under equal scan conditions. Custom creep functions can easily be added by the user.

\paragraph{interpolate}
This function removes the distortion induced by the sinusoidal tip trajectory. To this purpose, the sinusoidal movement is reconstructed and overlaid by a square grid. Hereby, the $x$-phase shift obtained in the \textit{assign frame} and, optionally, the creep parameters obtained in the \textit{determine creep} step are taken into account. The values associated with this fully described tip path are then used to interpolate onto a square grid. The Delaunay triangulation, described in Fig.\ \ref{fig:schematic}(c), serves as a basis for the high quality interpolation: the determined tip positions are divided into sets of three points that do not include further points within their circumcircles. The value of any point on the square grid is then calculated by a weighted sum of the three points surrounding it (normalized barycentric coordinates, see purple arrows in Fig.\ \ref{fig:schematic}(c). Importantly, this approach combines all corrections (except the drift) into a single interpolation step. Since the tip path is the same for all images, a significant acceleration of 4 orders of magnitude (in the limit of long movies) over readily available packages is achieved by performing the Delaunay triangulation only for a single image and then use it for the full 3D stack. This computation is performed in a matrix multiplication formalism, using sparse matrices to hold the barycentric interpolation coordinates.

\paragraph{correct drift}
The drift trajectory is calculated by a sliding FFT-based cross correlation (implemented with \texttt{scipy}) of frame pairs taken at a chosen time separation. In a first step, the frames are enlarged in the $x$- and $y$-dimensions to an appropriate power of 2 number of pixels. The time separation needs to be selected such that significant drift is detected between the compared frames, without losing track of drift details on shorter time scales. Drift values for the last frames for which no calculation is possible are linearly extrapolated from preceding ones. A median filter of variable width smoothes drift noise without removing abrupt changes in drift velocity. Additionally, an optional boxcar filter can be applied. The drift path is saved as a text file which can be adjusted manually and subsequently reloaded.\\
In the resulting movie, the drift is accounted for only by shifting entire frames by integer pixels since thermal drift within a single frame is typically minimal in fast acquired STM movies and the number of pixels can be increased accordingly. Two correction modes can be chosen for the drift corrected movie: It can be cropped down to the 'greatest common' area of all drifting frames, or, alternatively, the complete frames can be embedded into an enlarged array by zero padding ('full').

\paragraph{export} 
The final corrected movie is exported into an MPEG4 file or single frames into common image file formats or the Gwyddion native data format, \texttt{.gwy}, for further analysis \cite{dklapetek2012open}. Frame range, color map, contrast and frame rate can be adjusted and, optionally, frame numbers overlaid as text. It should be noted that the contrast is evaluated globally by a histogram, spanning over the entire frame range in an exported movie, allowing for quantitative comparisons of contrast (i.e. apparent heights) in different frames of the same movie.

\subsection{Sample code snippets analysis}

The movie conversion essentially consists of a few key steps which are summarized below. The respective steps in the workflow of Fig.\ \ref{fig:sketch} are indicated as code comments. The \textit{interpolate} step is divided into the construction of the Delaunay-based interpolation matrices and the actual interpolation. Having all parameters (see Supporting Information for an example of input parameters), these commands are sufficient to perform the complete movie conversion.

\begin{lstlisting}
import pyfastspm as pf

ft = pf.FastMovie(file_name)
ft.reshape_to_movie()  # assign frame
pf.filter_movie(ft.filterparam)  # FFT filter
creep = pf.Creep(ft, creep_mode='sin')  # determine creep
grid = creep.fit_creep()
matrix_up, matrix_down = pf.interpolate(
    ft,
    grid,
    give_grid=True,
)  # interpolate
pf.interpolate(ft, grid, matrix_up, matrix_down)
drift = pf.Drift(ft)  # correct drift
ft.data, drift_path = drift.correct(drift_type)
ft.export_movie
\end{lstlisting}
%\caption{Minimal sample script for movie conversion. The relation to the workflow steps displayed in Figure \ref{fig:sketch} is described in the code comments.}
%\label{lst:code_snippets}

\section{Illustrative Example}

\begin{figure}[htbp]
    \centering
    \includegraphics[width=0.99\textwidth]{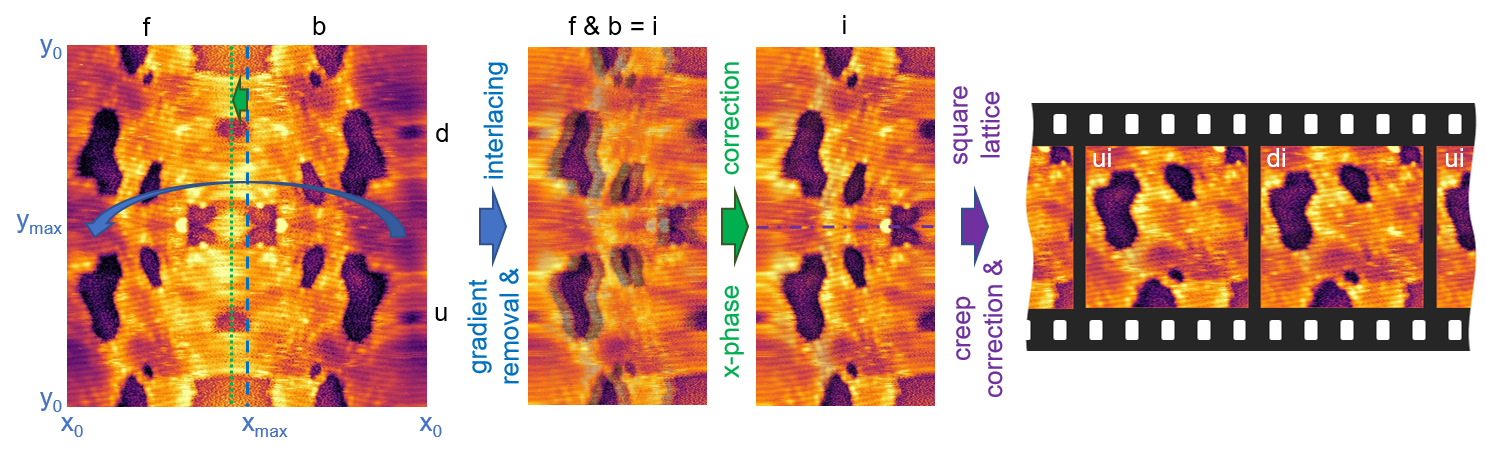}
    \caption{Effect of the conversion steps on an experimental example, a reduced Fe$_3$O$_4$(001) surface observed at \SI{657}{\kelvin} under \SI{5e-7}{\milli\bar} oxygen under which holes successively close, imaged at \SI{4.0}{\nano\ampere}, \SI{0.5}{\volt} and with an image size of $\SI{18}{nm} \times \SI{18}{nm}$. Movie frequencies: \SI{4}{fps} resp. \SI{2}{images/s}, fast $x$-frequency \SI{785}{\hertz}, pixel frequency \SI{307}{\kilo \hertz}. The left-most image contains the full information of one full scan cycle, including forward (f), backward (b), up (u) and down (d) motion. The tip moves in lines from left to right (fast $x$-direction), from bottom upwards and down again (slow $y$-direction). Each horizontal line in the shown image contains thus the forward and backward scan sequence, while the temporal succession from bottom upwards represents the various $y$-positions, first the upward, then the downward movement. The symmetry axes of this representation reveal the x- and $y$-phase shifts. 
    In a first step, FFT filtering removes the residual sample tilt gradient (typically strongest in the $x$-direction). To fold the f and b signals correctly onto each other into an interlaced (i) image, the correct $x$-phase shift (and thus the frame assignment of the pixels) has to be applied in a second step. Finally, the creep and sinusoidal distortions are removed, and the pixels interpolated onto a square lattice. The resulting two u and d frames can both contribute to a movie.}
    \label{fig:images}
\end{figure}

Fig.\ \ref{fig:images} shows the effect of the main conversion steps on a sample movie where the dynamic closing of holes is observed on a reduced magnetite Fe$_3$O$_4$(001) surface under an oxygen atmosphere at elevated temperatures. As described in detail in the figure caption, starting from the full image on the left, single movie frames on the right can be obtained, where every pixel is included without introducing flickering or distortions. The high quality of the image conversion is best observed in the Supporting Movie that displays the dynamics of a five-fold accelerated movie, comparing side-by-side the uncorrected (left) and corrected (right) data. The parameters used for this movie conversion are also included in the Supporting Information.

\begin{figure}[htbp]
    \centering
    \includegraphics{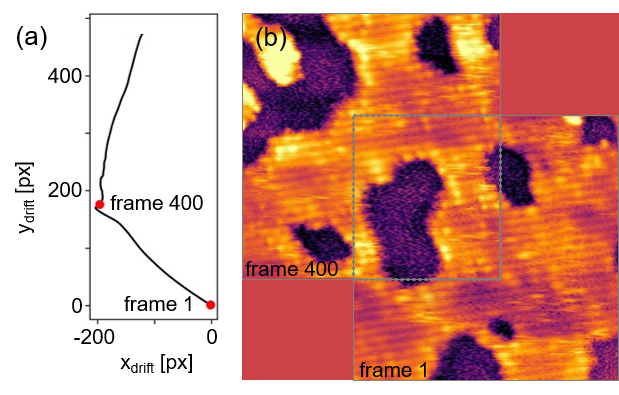}
    \caption{(a) Drift path of the movie shown in Fig. \ref{fig:images} as calculated by a sliding cross correlation of frame pairs taken at a chosen time separation (here \SI{20}{\second}; the drift for the last \SI{19}{\second} is thus extrapolated). (b) Illustration of the two possible export options on the example of two drift corrected frames. Either the full explored range is retained by padding with background pixels (all shown data) or the field of view is cropped to the greatest common area of all frames (gray dashed box).}
    \label{fig:drift}
\end{figure}

Once individual frames have been ideally reconstructed, thermal drift effects can be removed. As described above, the frames are simply shifted with respect to each other. This is done in a separate step that acts on the reconstructed 3D stack, illustrated in Fig. \ref{fig:drift}. The two alternative export modes, either cropping the 'greatest common' or exporting the 'full' area, are indicated in the figure. As the example shows, the drift correction algorithm is robust enough to cope not only with linear drift but also with discontinuities (here, for example around frame 400).

The full conversion and writing of this example movie file of 1044 frames can be performed fully automatically in a couple of minutes on a standard laptop (e.g. by using the example notebook available as part of the package), thus providing an easy and quick data analysis pathway.

\section{Impact}

\texttt{pyfastspm} provides a highly convenient and efficient tool for the non-specialized user to convert fast SPM data into publication quality movies. The software package thus paves the way for investigating surface dynamics that have typically only been accessible to a few specialized groups. In the four years of PyfastSPM development alone, the authors have demonstrated its power in three distinct examples of surface dynamics studies:
\begin{itemize}
    \item \textit{Imaging of short-lived intermediate species}, critical especially for the investigation of detailed mechanisms, e.g.\ in catalysis and thin film growth \cite{patera2018real}.
    \item \textit{Access to diffusion paths and statistics} beyond simple residence times, e.g.\ hopping of small metal clusters in pores of a 2D film revealing support symmetries \cite{lechner2018microscopy}. The improved image stability opens the possibility to quantitatively evaluate hopping processes with atomic resolution over large time intervals \cite{bourgund2019influence}.
    \item \textit{Recovery of atomic resolution} underneath highly mobile species which dominate the contrast in single SPM images (streaks), i.e.\ by averaging many drift-corrected frames \cite{arndt2020order}. Here, PyfastSPM provides a new modality for SPM imaging under extreme conditions (e.g.\ elevated temperatures, in gas atmospheres or at solid/liquid interfaces).
\end{itemize}

While all these features have already been feasible in principle, the FAST module \cite{dri2019new} and open source \texttt{pyfastspm} software package allow for a smooth switch between standard and time resolved SPM imaging, and a nearly perfect movie visualization close to real time. We expect that this will encourage users in the wider SPM community to routinely investigate surface dynamics, as was the experience in the daily practice of our own groups.

At the time of writing this article, the software package is in routine use in groups that are part of the European Nanoscience Foundries and Fine Analysis (NFFA) project. Further requests have already been received, from users that use the commercially available FAST module \cite{FASTwebsite}; the software package, however, is available open source. With appropriate minor modifications of the data import, \texttt{pyfastspm} empowers every SPM user to build publication ready movies from fast SPM data.

\section{Conclusions}

In conclusion, we have presented a freely available new open-source software module for easy and complete conversion of 1D fast SPM data streams into optimally corrected movie frames with the additional option of drift compensation. Specifically, these corrections concern noise reduction, background removal, phase shifts in the fast scanning direction, piezo creep distortions and removal of sinusoidal tip movement distortions. The result is a stable movie obtained without loss of any pixel information, which opens the way for the non-specialized user to routinely evaluate fast SPM data in a quantitative manner, down to the atomic scale. Examples were given for the investigation of elusive intermediate species, of fast diffusion dynamics, and of a new mode of image reconstruction by averaging under highly dynamic, extreme imaging conditions.

\section{Conflict of Interest}
No conflict of interest exists.

\section*{Acknowledgements}
\begin{figure}[h!]
    \includegraphics[width=0.1\textwidth]{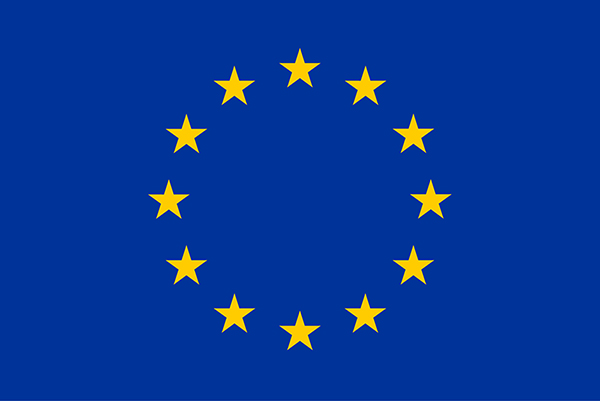}
\end{figure}
This project has received funding from the European Union's Horizon 2020 research and innovation programme under grant agreement No. 654360 and 101007417 within the framework of the NFFA-Europe and NFFA-Europe Pilot Joint Activities. The experimental work was funded by the Deutsche Forschungsgemeinschaft (DFG, German Research Foundation) under Germany's Excellence Strategy EXC 2089/1- 390776260 and project numbers ES 349/5-2 and ES 349/4-1). B.A.J.L. gratefully acknowledges financial support from the Young Academy of the Bavarian Academy of Sciences and Humanities. The authors would like to acknowledge the help from Cristina Africh (CNR-IOM) and contributions by Matthias Krinninger to the acquisition of the illustrative example movie.



%% References:
%\bibliographystyle{elsarticle-num} 
%\bibliography{bibliography}

\begin{thebibliography}{00}

\bibitem{henss2019density}
A.-K. Hen{\ss}, S.~Sakong, P.~K. Messer, J.~Wiechers, R.~Schuster, D.~C. Lamb,
  A.~Gro{\ss}, J.~Wintterlin, Density fluctuations as door-opener for diffusion
  on crowded surfaces, Science 363~(6428) (2019) 715--718.

\bibitem{bourgund2019influence}
A.~Bourgund, B.~A.~J. Lechner, M.~Meier, C.~Franchini, G.~S. Parkinson,
  U.~Heiz, F.~Esch, Influence of local defects on the dynamics of {O--H} bond
  breaking and formation on a magnetite surface, The Journal of Physical
  Chemistry C 123~(32) (2019) 19742--19747.

\bibitem{patera2018real}
L.~L. Patera, F.~Bianchini, C.~Africh, C.~Dri, G.~Soldano, M.~M. Mariscal,
  M.~Peressi, G.~Comelli, Real-time imaging of adatom-promoted graphene growth
  on nickel, Science 359~(6381) (2018) 1243--1246.

\bibitem{lechner2018microscopy}
B.~A.~J. Lechner, F.~Knoller, A.~Bourgund, U.~Heiz, F.~Esch, A microscopy
  approach to investigating the energetics of small supported metal clusters,
  The Journal of Physical Chemistry C 122~(39) (2018) 22569--22576.

\bibitem{wei2021electrochemical}
J.~Wei, Y.-X. Chen, O.~M. Magnussen, Electrochemical in situ video-{STM}
  studies of the phase transition of {CO} adlayers on {P}t(111) electrodes, The
  Journal of Physical Chemistry C 125~(5) (2021) 3066--3072.

\bibitem{rost2018high}
M.~Rost, High-speed electrochemical STM, Vol.~1, Elsevier, 2018.

\bibitem{schitter2008scanning}
G.~Schitter, M.~J. Rost, Scanning probe microscopy at video-rate, Materials
  Today 11 (2008) 40--48.

\bibitem{dri2019new}
C.~Dri, M.~Panighel, D.~Tiemann, L.~L. Patera, G.~Troiano, Y.~Fukamori,
  F.~Knoller, B.~A.~J. Lechner, G.~Cautero, D.~Giuressi, et~al., The new fast
  module: A portable and transparent add-on module for time-resolved
  investigations with commercial scanning probe microscopes, Ultramicroscopy
  205 (2019) 49--56.

\bibitem{Junkes:ICALEPCS2017-THPHA154}
H.~Junkes, H.-J. Freund, L.~Gura, M.~Heyde, P.~Marschalik, Z.~Yang,
  {E}xperiment
  {C}ontrol with {EPICS}7 and {S}ymmetric {M}ultiprocessing on {RTEMS}, in:
  Proc. of International Conference on Accelerator and Large Experimental
  Control Systems (ICALEPCS'17), Barcelona, Spain, 8-13 October 2017, no.~16 in
  International Conference on Accelerator and Large Experimental Control
  Systems, JACoW, Geneva, Switzerland, 2018, pp. 1762--1766.

\bibitem{computational_geometry}
M.~de~Berg, O.~Cheong, M.~van Kreveld, M.~Overmars, Computational Geometry:
  Algorithms and Applications, Springer Verlag, 2008, {ISBN:}
  978-91-637-4473-0.

\bibitem{arndt2020order}
B.~Arndt, B.~A.~J. Lechner, A.~Bourgund, E.~Gr{\aa}n{\"a}s, M.~Creutzburg,
  K.~Krausert, J.~Hulva, G.~S. Parkinson, M.~Schmid, V.~Vonk, F.~Esch,
  A.~Stierle, Order--disorder phase transition of the subsurface cation vacancy
  reconstruction on {F}e$_3${O}$_4$(001), Physical Chemistry Chemical Physics
  22~(16) (2020) 8336--8343.

\bibitem{rost2005scanning}
M.~Rost, L.~Crama, P.~Schakel, E.~Van~Tol, G.~van Velzen-Williams, C.~Overgauw,
  H.~Ter~Horst, H.~Dekker, B.~Okhuijsen, M.~Seynen, et~al., Scanning probe
  microscopes go video rate and beyond, Review of Scientific Instruments 76~(5)
  (2005) 053710.

\bibitem{choi2014growth}
J.~Choi, W.~Mayr-Schm{\"o}lzer, F.~Mittendorfer, J.~Redinger, U.~Diebold,
  M.~Schmid, The growth of ultra-thin zirconia films on {P}d$_3${Z}r(0001),
  Journal of Physics: Condensed Matter 26~(22) (2014) 225003.

\bibitem{dklapetek2012open}
P.~K. D~Ne{\v{c}}as, Gwyddion: An open-source software for {SPM} data analysis,
  Central European Journal of Physics 10 (2012) 181--188.

\bibitem{FASTwebsite}
Fast module webpage, \url{https://fastmodule.iom.cnr.it}.




\end{thebibliography}


% Please add the reference to the software repository if DOI for software  is available. 

% \section*{Supporting information}

\end{document}